\documentclass[titlepage,letterpaper,12pt]{article}
\usepackage[dvips]{graphicx}

\begin{document}

\title{Bounded solutions of neutral fermions with a screened Coulomb potential}
\date{}
\author{Antonio S. de Castro \\
\\
UNESP - Campus de Guaratinguet\'{a}\\
Departamento de F\'{\i}sica e Qu\'{\i}mica\\
Caixa Postal 205\\
12516-410 Guaratinguet\'{a} SP - Brasil\\
\\
E-mail address: castro@feg.unesp.br (A.S. de Castro)}
\date{}
\maketitle

\begin{abstract}
The intrinsically relativistic problem of a fermion subject to a
pseudoscalar screened Coulomb plus a uniform background potential in
two-dimensional space-time is mapped into a Sturm-Liouville. This mapping
gives rise to an effective Morse-like potential and exact bounded solutions
are found. It is shown that the uniform background potential determinates
the number of bound-state solutions. The behaviour of the eigenenergies as
well as of the upper and lower components of the Dirac spinor corresponding
to bounded solutions is discussed in detail and some unusual results are
revealed. An apparent paradox concerning the uncertainty principle is solved
by recurring to the concepts of effective mass and effective Compton
wavelength.
\end{abstract}

\section{Introduction}

The Coulomb potential of a point electric charge in a 1+1 dimension,
considered as the time component of a Lorentz vector, is linear ($\sim |x|$)
and so it provides a constant electric field always pointing to, or from,
the point charge. This problem is related to the confinement of fermions in
the Schwinger and in the massive Schwinger models \cite{col1}-\cite{col2}
and in the Thirring-Schwinger model \cite{fro}. It is frustrating that, due
to the tunneling effect (Klein\'{}s paradox), there are no bound states for
this kind of potential regardless of the strength of the potential \cite{cap}%
-\cite{gal}. The linear potential, considered as a Lorentz scalar, is also
related to the quarkonium model in one-plus-one dimensions \cite{hoo}-\cite
{kog}. Recently it was incorrectly concluded that even in this case there is
solely one bound state \cite{bha}. Later, the proper solutions for this last
problem were found \cite{cas1}-\cite{hil}. However, it is well known from
the quarkonium phenomenology in the real 3+1 dimensional world that the best
fit for meson spectroscopy is found for a convenient mixture of vector and
scalar potentials put by hand in the equations (see, e.g., \cite{luc}). The
same can be said about the treatment of the nuclear phenomena describing the
influence of the nuclear medium on the nucleons \cite{ser}-\cite{lis}. The
mixed vector-scalar potential has also been analyzed in 1+1 dimensions for a
linear potential \cite{cas2} as well as for a general potential which goes
to infinity as $|x|\rightarrow \infty $ \cite{ntd}. In both of those last
references it has been concluded that there is confinement if the scalar
coupling is of sufficient intensity compared to the vector coupling.
Although the vector Coulomb potential does not hold relativistic bound-state
solutions, its screened version ($\sim e^{-|x|/\lambda }$) is a genuine
binding potential and its solutions have been found for fermions \cite{ada}.
The screened Coulomb potential has also been analyzed with a scalar coupling
in the Dirac equation \cite{kag3} as well as in the Klein-Gordon equation
with vector \cite{kag1} and scalar \cite{kag2} couplings.

The confinement of fermions by a pseudoscalar double-step potential \cite
{asc2} and their scattering by a pseudoscalar step potential \cite{asc3}
have already been analyzed in the literature providing the opportunity to
find some quite interesting results. Indeed, the two-dimensional version of
the anomalous magnetic-like interaction linear in the radial coordinate,
christened by Moshinsky and Szczepaniak \cite{ms} as Dirac oscillator, has
also received attention. Nogami and Toyama \cite{nt}, Toyama et al. \cite
{tplus} and Toyama and Nogami \cite{tn} studied the behaviour of wave
packets under the influence of that conserving-parity potential whereas
Szmytkowski and Gruchowski \cite{sg} proved the completeness of the
eigenfunctions. More recently Pacheco et al. \cite{pa} studied some
thermodynamics properties of the 1+1 dimensional Dirac oscillator, and a
generalization of the Dirac oscillator for a negative coupling constant was
presented in Ref. \cite{asc}. The two-dimensional generalized Dirac
oscillator plus an inversely linear potential has also been addressed \cite
{asc4}. In  recent papers, Villalba \cite{vil} and McKeon and Van Leeuwen
\cite{mck} considered a pseu\-do\-sca\-lar Coulomb potential ($V=\lambda /r$%
) in 3+1 dimensions and concluded that there are no bounded solutions. The
reason attributed in Ref. \cite{mck} for the absence of bounded solutions it
that the different parity eigenstates mix. Furthermore, the authors of Ref.
\cite{mck} assert that \textit{the absence of bound states in this system
confuses the role of the }$\pi $\textit{-meson in the binding of nucleons}.
Such an intriguing conclusion sets the stage for the analyses by other sorts
of pseudoscalar potentials. A natural question to ask is if the absence of
bounded solutions by a pseudoscalar Coulomb potential is a characteristic
feature of the four-dimensional world. In Ref. \cite{asc} the Dirac equation
in one-plus-one dimensions with the pseudoscalar power-law potential $V=\mu
|x|^{\delta }$ was approached and there it was concluded that only for $%
\delta >0$ there can be a binding potential. Furthermore, we succeed in
searching bound-state solutions for the pseudoscalar Coulomb potential ($%
V=m\omega |x|$) as well as for the pseudoscalar potential $V=m\omega x$,
even in the case $\omega <0$ \cite{asc}.

In the present paper we begin our presentation of the Dirac equation with
the most general potential in 1+1 dimensions. This general approach allows
us to show that the pseudoscalar potential has a clearly different behaviour
when compared with the scalar and vector potentials, even in the
nonrelativistic limit of the theory. It was shown in Ref. \cite{asc4} that
the problem of a fermion under the influence of a pseudoscalar potential can
always be mapped into a Sturm-Liouville problem for the upper and lower
components of the Dirac spinor. Following that same methodology the present
paper shows that the screened Coulomb potential plus a uniform background
potential gives rise to an effective Morse-like potential, in the same way
as the generalized Dirac oscillator plus an inversely linear potential gives
rise to an effective quadratic plus inversely quadratic potential \cite{asc4}%
. The exact bounded solutions are found, such as in Ref. \cite{asc4}, by
transforming the Sturm-Liouville problem into Kummer's equation and
expressing the upper and the lower components of the Dirac spinor in terms
of confluent hypergeometric functions. In addition to their intrinsic
importance as a new solution of the Dirac equation, this problem highlights
the essential role of the uniform background potential in furnishing bounded
solutions and it might be relevant to studies of binding of neutral fermions
by electric fields. Furthermore, the result renders a new contrast to the
result found in Ref. \cite{mck}. In the nonrelativistic quantum mechanics,
the problem of a particle subject to the Morse potential ($V=V_{0}\left[
1-\exp (-ar)\right] ^{2}$) has been used to describe the vibrations of
nuclei in homonuclear diatomic molecules \cite{mor1}-\cite{mor2}. This sort
of problem for S-wave states is transformed into the problem of solving a
transcendent equation which is only approximately solvable \cite{flu}-\cite
{haa1}. Nevertheless, the one-dimensional asymmetric Morse potential ($%
-\infty <r<\infty $) is an exactly solvable problem in the nonrelativistic
quantum mechanics \cite{haa}-\cite{nie}, even if its parameters are complex
numbers \cite{ahmed}.

\section{The Dirac equation in a 1+1 dimension}

Thee 1+1 dimensional time-independent Dirac equation for a fermion of rest
mass $m$ reads
\begin{equation}
\mathcal{H}\Psi =E\Psi ,\quad \mathcal{H}=c\alpha p+\beta mc^{2}+\mathcal{V}
\label{eq1}
\end{equation}

\noindent where $E$ is the energy of the fermion, $c$ is the velocity of
light and $p$ is the momentum operator. We use $\alpha =\sigma _{1}$ and $%
\beta =\sigma _{3}$, where $\sigma _{1}$ and $\sigma _{3}$ are Pauli
matrices. For the potential matrix we consider
\begin{equation}
\mathcal{V}=1V_{t}+\beta V_{s}+\alpha V_{e}+\beta \gamma ^{5}V_{p}
\label{eq2}
\end{equation}

\noindent where $1$ stands for the 2$\times $2 identity matrix and $\beta
\gamma ^{5}=\sigma _{2}$. This is the most general combination of Lorentz
structures for the potential matrix because there are only four linearly
independent 2$\times $2 matrices. The subscripts for the terms of potential
denote their properties under a Lorentz transformation: $t$ and $e$ for the
time and space components of the 2-vector potential, $s$ and $p$ for the
scalar and pseudoscalar terms, respectively. It is worth to note that the
Dirac equation is covariant under $x\rightarrow -x$ if $V_{e}(x)$ and $%
V_{p}(x)$ change sign whereas $V_{t}(x)$ and $V_{s}(x)$ remain the same.
This is because the parity operator $P=\exp (i\eta )P_{0}\sigma _{3}$, where
$\eta $ is a constant phase and $P_{0}$ changes $x$ into $-x$, changes sign
of $\alpha $ and $\beta \gamma ^{5}$ but not of $1$ and $\beta $.

Defining
\begin{equation}
\psi =\exp \left( \frac{i}{\hbar }\Lambda \right) \Psi ,\quad \Lambda
(x)=\int^{x}dx^{\prime }\frac{V_{e}(x^{\prime })}{c}  \label{eq5}
\end{equation}

\noindent the space component of the vector potential is gauged away

\begin{equation}
\left( p+\frac{V_{e}}{c}\right) \Psi =\exp \left( \frac{i}{\hbar }\Lambda
\right) p\psi  \label{eq7}
\end{equation}

\noindent so that the time-independent Dirac equation can be rewritten as
follows:

\begin{equation}
H\psi =E\psi ,\quad H=\sigma _{1}cp+\sigma _{2}V_{p}+\sigma _{3}\left(
mc^{2}+V_{s}\right) +1V_{t}  \label{eq7a}
\end{equation}

\noindent showing that the space component of a vector potential only
contributes to change the spinors by a local phase factor.

Provided that the spinor is written in terms of the upper and the lower
components
\begin{equation}
\psi =\left(
\begin{array}{c}
\psi _{+} \\
\psi _{-}
\end{array}
\right)  \label{eq8a}
\end{equation}

\noindent the Dirac equation decomposes into:

\begin{equation}
\left( V_{t}-E\pm V_{s}\pm mc^{2}\right) \psi _{\pm }=i\hbar c\psi _{\mp
}^{\prime }\pm iV_{p}\psi _{\mp }  \label{eq8b}
\end{equation}

\noindent where the prime denotes differentiation with respect to $x$. In
terms of $\psi _{+}$ and $\psi _{-}$ the spinor is normalized as $%
\int_{-\infty }^{+\infty }dx\left( |\psi _{+}|^{2}+|\psi _{-}|^{2}\right) =1$
so that $\psi _{+}$ and $\psi _{-}$ are square integrable functions. It is
clear from the pair of coupled first-order differential equations (\ref{eq8b}%
) that both $\psi _{+}$ and $\psi _{-}$ have definite and opposite parities
if the Dirac equation is covariant under $x\rightarrow -x$.

In the nonrelativistic approximation (potential energies small compared to $%
mc^{2}$ and $E\approx mc^{2}$) Eq. (\ref{eq8b}) becomes

\begin{equation}
\psi _{-}=\frac{p}{2mc}\,\psi _{+}  \label{eq8c}
\end{equation}

\begin{equation}
\left( -\frac{\hbar ^{2}}{2m}\frac{d^{2}}{dx^{2}}+V_{t}+V_{s}+\frac{V_{p}^{2}%
}{2mc^{2}}+\frac{\hbar V_{p}^{\prime }}{2mc}\right) \psi _{+}=\left(
E-mc^{2}\right) \psi _{+}  \label{eq8d}
\end{equation}

\noindent Eq. (\ref{eq8c}) shows that $\psi _{-}$ is of order $v/c<<1$
relative to $\psi _{+}$ and Eq. (\ref{eq8d}) shows that $\psi _{+}$ obeys
the Schr\"{o}dinger equation without distinguishing the contributions of
vector and scalar potentials (at this point the author digress to make his
apologies for mentioning in former papers (\cite{asc2}-\cite{asc3}) that the
pseudoscalar potential does not present any contributions in the
nonrelativistic limit).

It is remarkable that the Dirac equation with a nonvector potential, or a
vector potential contaminated with some scalar or pseudoscalar coupling, is
not invariant under $V\rightarrow V+$const, this is so because only the
vector potential couples to the positive-energies in the same way it couples
to the negative-ones, whereas nonvector contaminants couple to the mass of
the fermion. Therefore, if there is any nonvector coupling the absolute
values of the energy will have physical significance and the freedom to
choose a zero-energy will be lost. This last statement remains truthfully in
the nonrelativistic limit if one considers that such a contaminant is a
pseudoscalar potential. It is also noticeable that the pseudoscalar coupling
results in the Schr\"{o}dinger equation with an effective potential in the
nonrelativistic limit, and not with the original potential itself. Indeed,
this is the side effect which in a 3+1 dimensional space-time makes the
linear potential to manifest itself as a harmonic oscillator plus a strong
spin-orbit coupling in the nonrelativistic limit \cite{ms}. The form in
which the original potential appears in the effective potential, the $%
V_{p}^{2}$ term, allows us to infer that even a potential unbounded from
below could be a confining potential. This phenomenon is inconceivable if
one starts with the original potential in the nonrelativistic equation.

As we will see explicitly in the next section, a constant added to the
pseudoscalar Coulomb potential is undoubtedly physically relevant. As a
matter of fact, it plays a crucial role to ensure the existence of bounded
solutions. Nevertheless, the resulting potential does not present any
nonrelativistic limit. Oddly enough, $E\approx mc^{2}$ only for the
positive-ground-state solution in the strong-coupling regime of the theory.

\section{The pseudoscalar screened Coulomb potential}

From now on we shall restrict our discussion to a pure pseudoscalar
potential. For $E\neq \pm mc^{2}$, the coupling between the upper and the
lower components of the Dirac spinor can be formally eliminated when Eq. (%
\ref{eq8b}) is written as second-order differential equations:

\begin{equation}
-\frac{\hbar ^{2}}{2m}\;\psi _{\pm }^{\prime \prime }+V_{eff}^{\left[ \pm
\right] }\;\psi _{\pm }=E_{eff}\;\psi _{\pm }  \label{30}
\end{equation}

\noindent where
\begin{eqnarray}
E_{eff} &=&\frac{E^{2}-m^{2}c^{4}}{2mc^{2}}  \label{30-1} \\
&&  \nonumber \\
V_{eff}^{\left[ \pm \right] } &=&\frac{V^{2}}{2mc^{2}}\pm \frac{\hbar }{2mc}%
V^{\prime }  \label{30-2}
\end{eqnarray}

\bigskip \noindent \noindent \noindent \noindent These last results show
that the solution for this class of problem consists in searching for
bounded solutions for two Schr\"{o}dinger equations. It should not be
forgotten, though, that the equations for $\psi _{+}$ or $\psi _{-}$ are not
indeed independent because the effective eigenvalue, $E_{eff}$, appears in
both equations. Therefore, one has to search for bound-state solutions for $%
V_{eff}^{\left[ +\right] }$ and $V_{eff}^{\left[ -\right] }$ with a common
eigenvalue. The Dirac eigenvalues are obtained by inserting the effective
eigenvalues in (\ref{30-1}).

The solutions for $E=\pm mc^{2}$, excluded from the Sturm-Liouville problem,
can be obtained directly from the Dirac equation (\ref{eq8b}). One can
observe that such a sort of isolated solutions can be written as

\begin{eqnarray}
\psi _{\mp } &=&N_{\mp }\,\exp \left[ \mp v(x)\right]  \nonumber \\
&&  \label{30-3} \\
\psi _{\pm }^{\prime }\mp v^{\prime }\psi _{\pm } &=&\pm i\,\frac{2mc}{\hbar
}N_{\mp }\,\exp \left[ \mp v(x)\right]  \nonumber
\end{eqnarray}

\noindent where $N_{+}$ and $N_{-}$ are normalization constants and $%
v(x)=\int^{x}dy\,V(y)\,/(\hbar c)$. \noindent One can check that it is
impossible to have both components different from zero simultaneously on the
same side of the $x$-axis. Of course normalizable eigenstates are possible
only if $v(x)$ has a distinctive leading asymptotic behaviour.

Now let us focus our attention on a pseudoscalar screened Coulomb potential
in the form
\begin{equation}
V=\frac{\hbar c}{2\lambda }\left[ g_{0}-g\exp \left( -\frac{|x|}{\lambda }%
\right) \right]  \label{30b}
\end{equation}
\noindent where the coupling constants, $g_{0}$ and $g$, are real numbers
and $\lambda $ is a positive parameter related to the range of the
interaction. Although $g<0$ gives rise to an ubiquitous repulsive potential
in a nonrelativistic theory, the possibility of such a sort of potential to
bind fermions is already noticeable in the nonrelativistic limit of the
Dirac theory (see Eq. (\ref{eq8d})). The presence of the uniform background
potential, $\hbar cg_{0}/(2\lambda )$, is a distinguishing feature from the
screened Coulomb potential in Refs. \cite{ada}-\cite{kag2}. As one will see,
it can not vanish for a binding pseudoscalar potential, and there is a
minimum size for the coupling constant $g_{0}$ before any bounded solution
can be obtained, viz. $|g_{0}|=1$. Evidently, the bound-state solutions for
this potential present no nonrelativistic limit, hence we have to cope with
an intrinsically relativistic bound-state problem.

We begin with the isolated solutions. In this case
\begin{equation}
v(x)=\frac{g_{0}}{2\lambda }\,x+\varepsilon (x)\frac{g}{2}\exp \left( -\frac{%
|x|}{\lambda }\right)  \label{30c}
\end{equation}

\noindent where $\varepsilon (x)$ stands for the sign function ($\varepsilon
(x)=x/|x|$ for $x\neq 0$). It is not difficult to see that $\psi _{+}$ and $%
\psi _{-}$ are not continuous at $x=0$, and for that reason those solutions
should be discarded.

As for $E\neq 0$, the effective potential becomes the Morse-like potential
\begin{equation}
V_{eff}^{\left[ \pm \right] }=A\exp \left( -2\frac{|x|}{\lambda }\right)
+B_{\pm }\,\exp \left( -\frac{|x|}{\lambda }\right) +C  \label{13}
\end{equation}
\noindent where
\begin{equation}
A=\frac{\hbar ^{2}g^{2}}{8\lambda ^{2}m}>0,\quad B_{\pm }=-\,\frac{\hbar
^{2}g}{4\lambda ^{2}m}\left[ g_{0}\mp \varepsilon (x)\right] ,\quad C=\frac{%
\hbar ^{2}g_{0}^{2}}{8\lambda ^{2}m}>0  \label{14}
\end{equation}

Before proceeding, it is useful to make some qualitative arguments regarding
the Morse-like potential and its possible solutions. The effective potential
is able to bind fermions on the condition that $B_{\pm }<0$, contrariwise $%
V_{eff}^{\left[ +\right] }$, or $V_{eff}^{\left[ -\right] }$, or both of
them would be repulsive everywhere. It follows that $\varepsilon (g)g_{0}>1$%
. It also follows that $V_{\min }^{\left[ \pm \right] }<E_{eff}<C$, where $%
V_{\min }^{\left[ \pm \right] }$ is the lowest value of $V_{eff}^{\left[ \pm
\right] }$. This implies that the Dirac eigenvalues corresponding to bounded
solutions are in the range $m^{2}c^{4}<E^{2}<m^{2}c^{4}+2mc^{2}C$. There is
a spectral gap in the range $E^{2}<m^{2}c^{4}$, and the energies belonging
to $E^{2}>m^{2}c^{4}+2mc^{2}C$ correspond to the continuum.

Note that the range of discrete Dirac eigenvalues enlarges as $\lambda $
decreases or $|g_{0}|$ increases. As a matter of fact, when $\lambda $
decreases the potential well becomes deeper and beyond this it becomes
narrower in such a way that $\lambda ^{2}V_{\min }^{\left[ \pm \right] }$ is
approximately constant, thus the capacity of the potential well to hold
bound-state solutions remains almost the same. As for $|g_{0}|$, its
increasing makes the potential well deeper almost without modifying its
width. In this way, one expects that the parameter $g_{0}$ determinates the
number of allowed bounded solutions.

The Morse-like potential presents, for $B_{\pm }<0$, a structure of
asymmetric potential wells and the highest well governs the value of the
zero-point energy. When $g\rightarrow 0$, $V_{eff}^{\left[ \pm \right] }$
has a single-well structure and nondegenerate Dirac eigenvalues are
expected. When $g\rightarrow \infty $, though, there appears a double-well
structure with an infinitely high and broad potential wall at the origin
separating the two wells. In this last circumstance one would expect
doubly-degenerate energy levels.

Note that the parameters of the effective potential are related in such a
manner that the change $x\rightarrow -x$ induces the change $V_{eff}^{\left[
\pm \right] }\rightarrow V_{eff}^{\left[ \mp \right] }$ without affecting
the effective energy, signifying that $|\psi _{\pm }(-x)|$ behaves like $%
|\psi _{\mp }(x)|$ and permitting us to focus our attention for a while on
the positive side of the $x$-axis.

Now let us do it quantitatively. Defining the dimensionless quantities $z$, $%
\mu _{\varepsilon (g)}^{\left[ \pm \right] }$ and $\nu $,

\negthinspace
\[
z=|g|\exp \left( -\frac{x}{\lambda }\right) \qquad
\]

\begin{equation}
\mu _{\varepsilon (g)}^{\left[ \pm \right] }=\frac{\varepsilon (g)}{2}\left(
g_{0}\mp 1\right)  \label{15b}
\end{equation}

\[
\nu =\frac{\lambda }{\hbar c}\sqrt{\left( \frac{\hbar cg_{0}}{2\lambda }%
\right) ^{2}+m^{2}c^{4}-E^{2}}
\]

\noindent and using (\ref{30})-(\ref{30-1}) and (\ref{13})-(\ref{14}), one
obtains the equation on the positive side of the $x$-axis

\begin{equation}
z\,\psi _{\pm }^{\prime \prime }+\psi _{\pm }^{\prime }+\left( -\frac{z}{4}-%
\frac{\nu ^{2}}{z}+\mu _{\varepsilon (g)}^{\left[ \pm \right] }\right) \psi
_{\pm }=0  \label{16}
\end{equation}

\noindent Now the prime denotes differentiation with respect to $z$.
Following the steps of Refs. \cite{kag1} and \cite{kag2}, we make the
transformation $\psi _{\pm }=z^{-1/2}\Phi _{\pm }$ to obtain the Whittaker
equation \cite{abr}:
\begin{equation}
\Phi _{\pm }^{\prime \prime }+\left( -\frac{1}{4}+\frac{\mu _{\varepsilon
(g)}^{\left[ \pm \right] }}{z}+\frac{1/4-\nu ^{2}}{z^{2}}\right) \Phi _{\pm
}=0  \label{16a}
\end{equation}
whose solution vanishing at the infinity becomes
\begin{equation}
\Phi _{\pm }=N_{\pm }\,z^{\nu +1/2}e^{-z/2}M\left( a_{\varepsilon
(g)}^{\left[ \pm \right] },b,z\right)  \label{16b}
\end{equation}
\noindent where $N_{+}$ and $N_{-}$ are constants, and $M$ is the regular
solution of the confluent hypergeometric equation (Kummer\'{}s equation)
\cite{abr}:

\begin{equation}
zM^{\prime \prime }+(b-z)M^{\prime }-a_{\varepsilon (g)}^{\left[ \pm \right]
}M=0  \label{19}
\end{equation}

\noindent with

\begin{equation}
a_{\varepsilon (g)}^{\left[ \pm \right] }=\nu +\frac{1}{2}-\frac{\varepsilon
(g)}{2}\left( g_{0}\mp 1\right) ,\qquad b=2\nu +1  \label{19a}
\end{equation}
\noindent Now we are ready to write the physically acceptable solutions on
both sides of the $x$-axis by recurring to the symmetry $|\psi _{\pm
}(-x)|\sim |\psi _{\mp }(x)|$ mentioned before. They are

\begin{eqnarray}
\psi _{+} &=&z^{\nu }e^{-z/2}\left[ \theta (-x)C^{\left[ -\right] }M\left(
a_{\varepsilon (g)}^{\left[ -\right] },b,z\right) +\theta (+x)C^{\left[
+\right] }M\left( a_{\varepsilon (g)}^{\left[ +\right] },b,z\right) \right]
\nonumber \\
&&  \label{eq10} \\
\psi _{-} &=&z^{\mu }e^{-z/2}\left[ \theta (-x)D^{\left[ -\right] }M\left(
a_{\varepsilon (g)}^{\left[ +\right] },b,z\right) +\theta (+x)D^{\left[
+\right] }M\left( a_{\varepsilon (g)}^{\left[ -\right] },b,z\right) \right]
\nonumber
\end{eqnarray}

\noindent where $C^{\left[ \pm \right] }$ and $D^{\left[ \pm \right] }$ are
constants, and $\theta (x)$ is the Heaviside function.

The continuity of the wavefunctions (\ref{eq10}) at $x=0$ furnishes
\begin{equation}
C^{\left[ +\right] }D^{\left[ +\right] }=C^{\left[ -\right] }D^{\left[
-\right] }  \label{10-1}
\end{equation}

\begin{equation}
\frac{C^{\left[ +\right] }}{D^{\left( +\right) }}=\frac{C^{\left[ -\right] }%
}{D^{\left[ -\right] }}\left[ \frac{M\left( a_{\varepsilon (g)}^{\left[
+\right] }-\varepsilon (g),b,z_{0}\right) }{M\left( a_{\varepsilon
(g)}^{\left[ +\right] },b,z_{0}\right) }\right] ^{2}  \label{22b}
\end{equation}
where $z_{0}=|g|$. Substituting the solutions (\ref{eq10}) into the Dirac
equation (\ref{eq8b}) and making use of the recurrence formulas \cite{abr}
\[
M^{\prime }(a,b,z)=\frac{a}{b}M(a+1,b+1,z)
\]
\begin{equation}  \label{22}
\end{equation}
\[
bM(a,b,z)-bM(a-1,b,z)-zM(a,b+1,z)=0
\]

\noindent one has as a result

\begin{equation}
\frac{C^{\left[ \pm \right] }}{D^{\left[ \pm \right] }}=\pm \left[ \frac{%
i\hbar c}{\lambda }\,\frac{a_{\varepsilon (g)}^{\left[ -\varepsilon
(g)\right] }}{E\mp \varepsilon (g)mc^{2}}\right] ^{\pm \varepsilon (g)}
\label{22a}
\end{equation}

\noindent

\noindent Together, (\ref{22b}) and (\ref{22a}) lead to the quantization
condition

\begin{equation}
\left[ \frac{M\left( a_{\varepsilon (g)}^{\left[ +\right] }-\varepsilon
(g),b,z_{0}\right) }{M\left( a_{\varepsilon (g)}^{\left[ +\right]
},b,z_{0}\right) }\right] ^{2}+1-\frac{g_{0}}{a_{-}^{\left[ +\right] }}=0
\label{eq23}
\end{equation}

\begin{figure}[!ht]
\begin{center}
\includegraphics[width=8cm, angle=270]{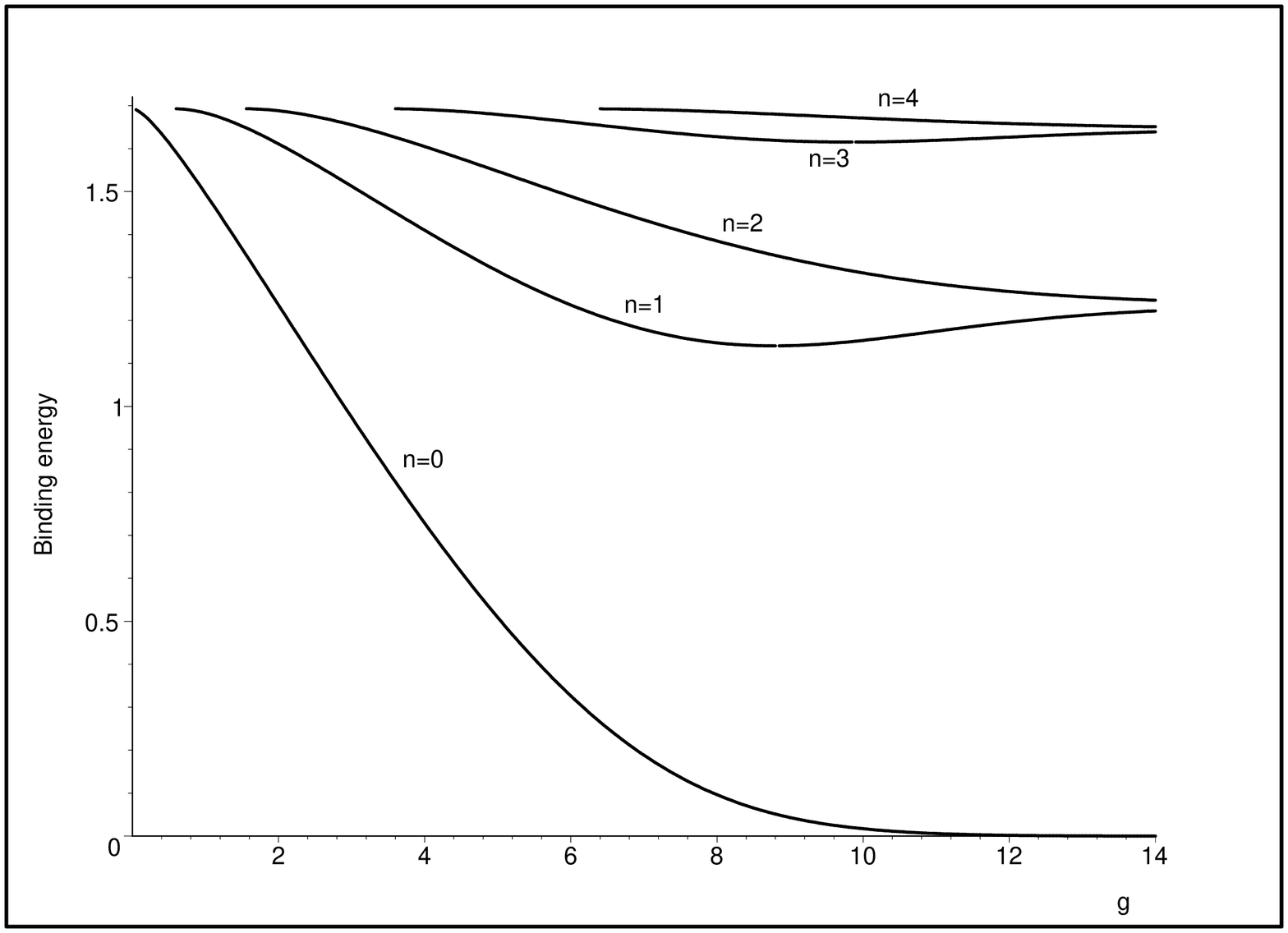}
\end{center}
\par
\vspace*{-0.1cm}
\caption{Binding energies ($E-mc^{2}$) corresponding to positive Dirac
eigenvalues as a function of $g$ with $\lambda =\hbar/(mc)$ and $g_{0}=5$ ($%
m=c=\hbar =1$).}
\label{Fig1}
\end{figure}

\noindent This last equation only holds for $g_{0}/a_{-}^{\left[ +\right]
}>1 $, and it represents a limit imposed on $E^{2}$; namely, $%
E^{2}>m^{2}c^{4}$, as has been anticipated by the preceding qualitative
arguments. The transcendental eigenvalue equation (\ref{eq23}) is invariant
under the transformation $E\rightarrow -E$ as well as under the combined
transformations $g_{0}\rightarrow -g_{0}$ and $g\rightarrow -g$, and it can
be solved easily with a symbolic algebra program by searching for solutions
in the range $m^{2}c^{4}<E^{2}<m^{2}c^{4}+(\hbar cg_{0}/2\lambda )^{2}$,
with $\varepsilon (g)g_{0}>1$. Fig. \ref{Fig1} shows the behaviour of the
positive-eigenenergies as a function of $g$ for $\lambda $ equal to the
Compton wavelength, i.e. $\lambda =\hbar /(mc)$, and $g_{0}=5$. There is a
finite number of eigenvalues. Notice that for very small values of $g$, at
least one energy level emerging from the continuum comes into existence. The
ground state tends asymptotically to $\pm mc^{2}$ for arbitrary large $g$
whereas the excited states tend to be closely bunched in pairs with
eigenvalues independent of $g$. Due to the appearance of the potential
barrier for large $g$, this sort of two-fold degeneracy comes as no surprise.

\begin{figure}[!ht]
\begin{center}
\includegraphics[width=8cm, angle=270]{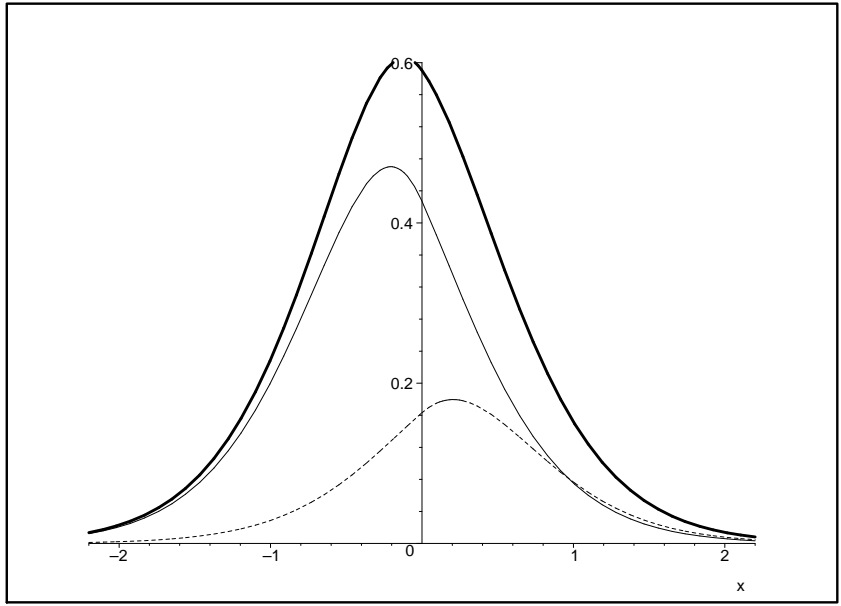}
\end{center}
\par
\vspace*{-0.1cm}
\caption{$|\psi _{+}|^{2}$ (full thin line), $|\psi _{-}|^{2}$ (dashed line)
and $|\psi |^{2}=|\psi _{+}|^{2}+|\psi _{-}|^{2}$ (full thick line) as a
function of $x$, corresponding to the positive-ground-state energy ($n=0$)
with $\lambda =\hbar/(mc)$, $g=2$ and $g_{0}=5$ ($m=c=\hbar =1$).}
\label{Fig2}
\end{figure}

\begin{figure}[!ht]
\begin{center}
\includegraphics[width=8cm, angle=270]{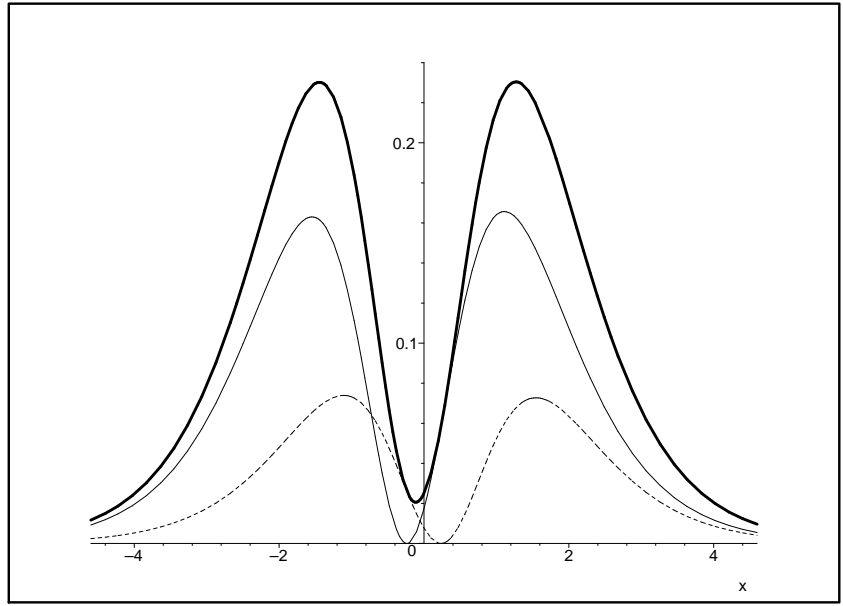}
\end{center}
\par
\vspace*{-0.1cm}
\caption{The same as in Fig. 2 corresponding to positive-first-excited-state
energy ($n=1$).}
\label{Fig3}
\end{figure}

\begin{figure}[!ht]
\begin{center}
\includegraphics[width=8cm, angle=270]{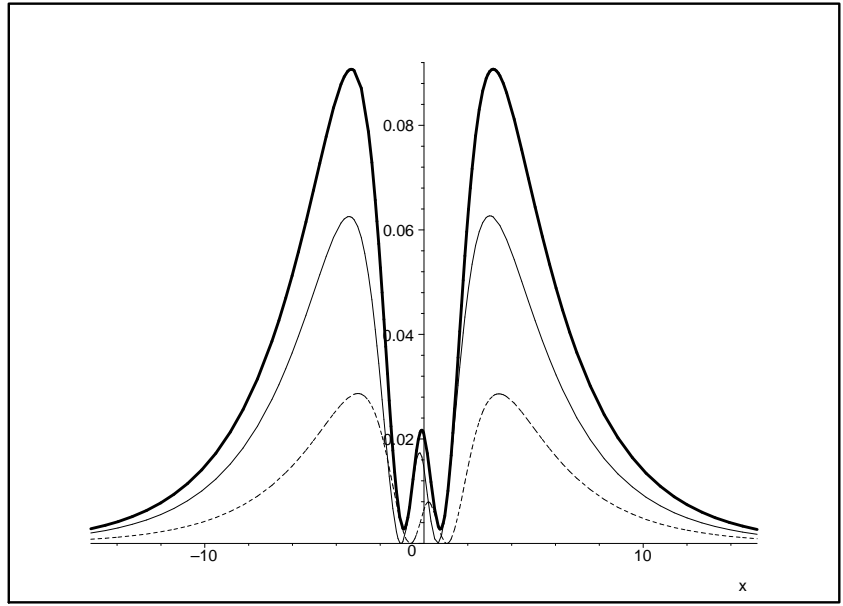}
\end{center}
\par
\vspace*{-0.1cm}
\caption{The same as in Fig. 2 corresponding to
positive-second-excited-state energy ($n=2$).}
\label{Fig4}
\end{figure}

\begin{figure}[!ht]
\begin{center}
\includegraphics[width=8cm, angle=270]{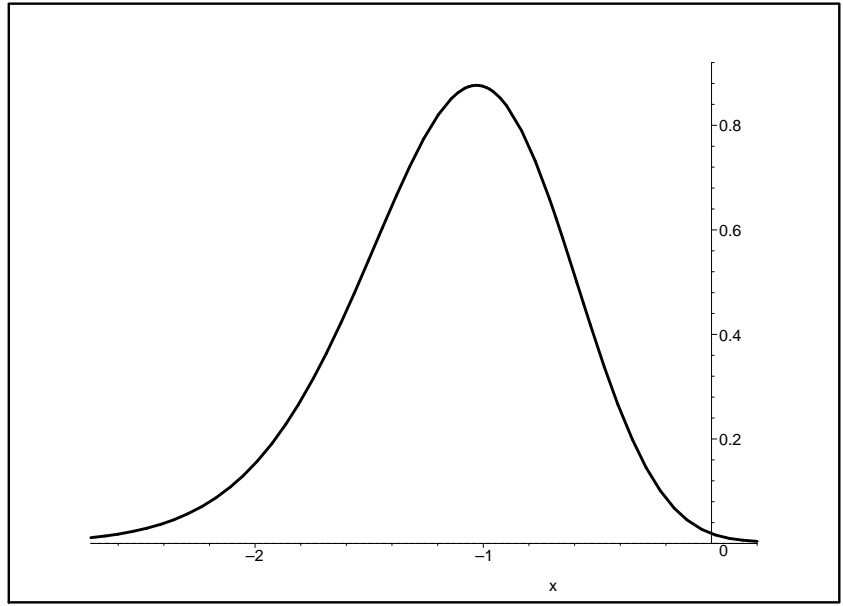}
\end{center}
\par
\vspace*{-0.1cm}
\caption{The same as in Fig. 2 for $g=14$. Here $|\psi _{-}|^{2}\ll |\psi
_{+}|^{2}$, hence $|\psi |^{2}\approx|\psi _{+}|^{2}$. }
\label{Fig5}
\end{figure}

\begin{figure}[!ht]
\begin{center}
\includegraphics[width=8cm, angle=270]{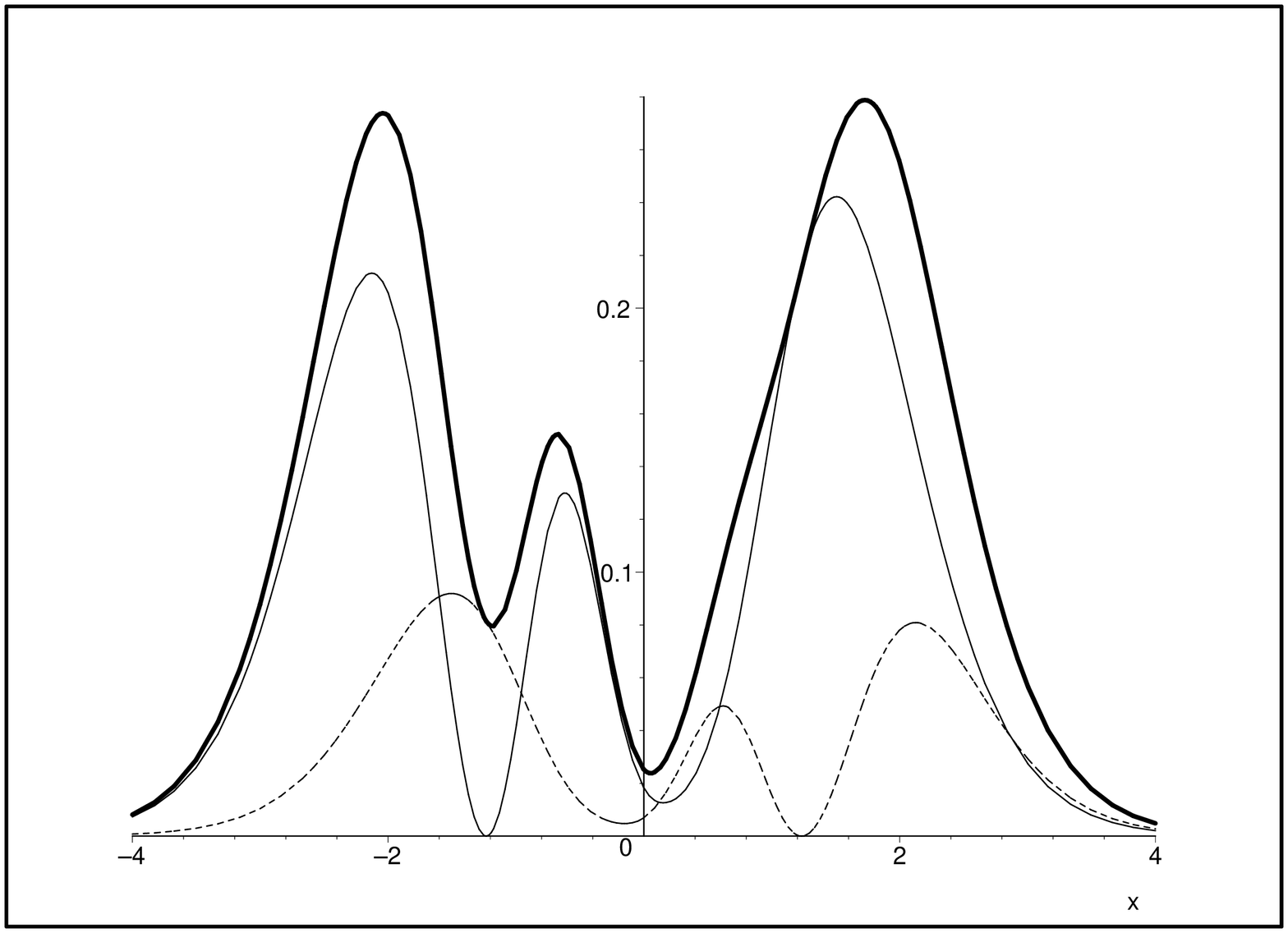}
\end{center}
\par
\vspace*{-0.1cm}
\caption{The same as in Fig. 2 for $n=1$ and $g=14$.}
\label{Fig6}
\end{figure}

\begin{figure}[!ht]
\begin{center}
\includegraphics[width=8cm, angle=270]{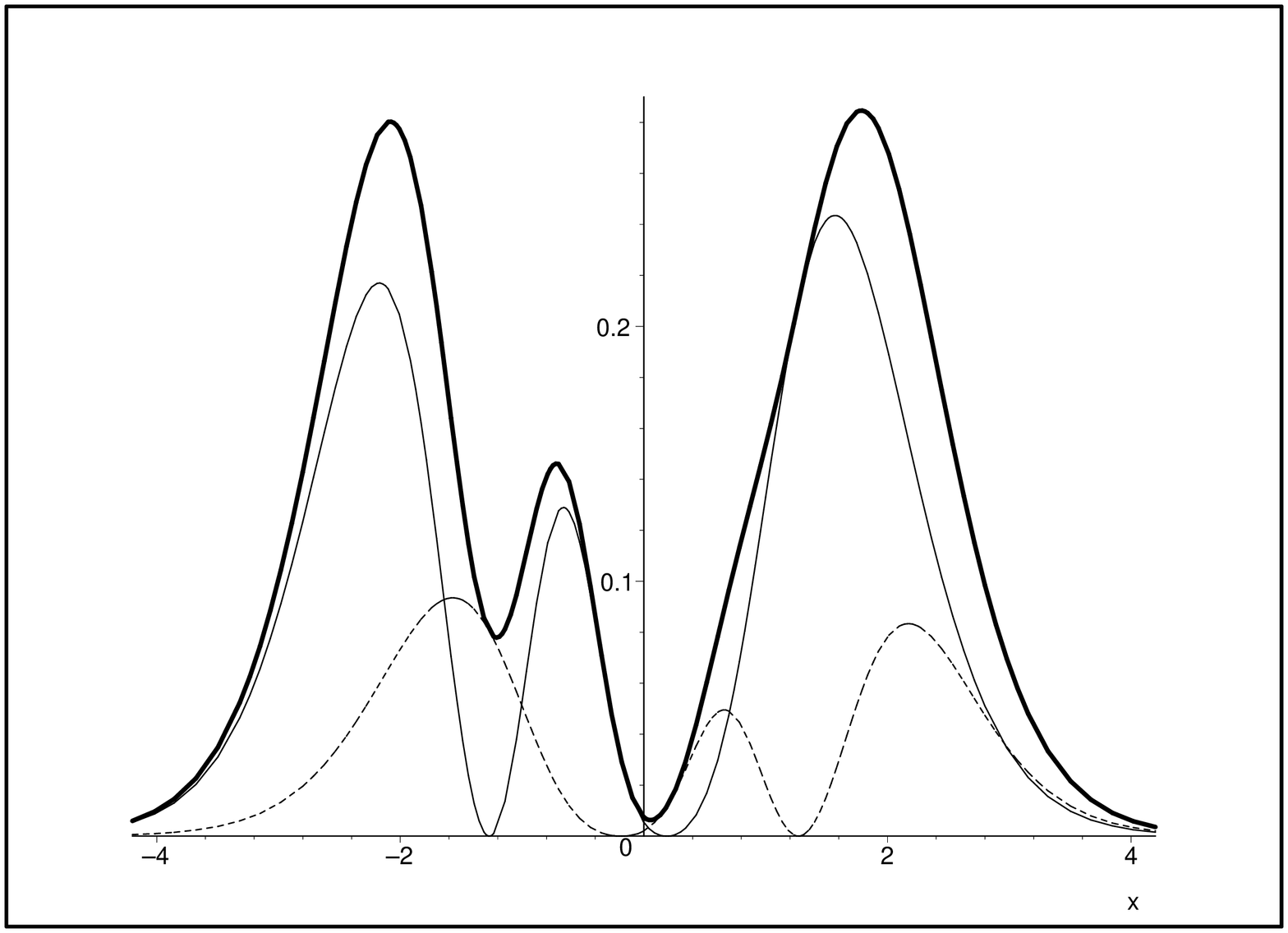}
\end{center}
\par
\vspace*{-0.1cm}
\caption{The same as in Fig. 2 for $n=2$ and $g=14$.}
\label{Fig7}
\end{figure}

Now we return our attention to (\ref{eq10}), where it remains to evaluate
the constants $C^{\left[ \pm \right] }$ and $D^{\left[ \pm \right] }$. Using
(\ref{10-1}) and (\ref{22a}) one obtains

\[
D^{\left[ +\right] }=\left[ \frac{i\hbar c}{\lambda }\,\frac{a_{-\varepsilon
(g)}^{\left[ \varepsilon (g)\right] }}{E+\varepsilon (g)mc^{2}}\right]
^{\varepsilon (g)}\,C^{\left[ +\right] }
\]

\begin{equation}
C^{\left[ -\right] }=\left[ \frac{\hbar c}{\lambda }\,\frac{a_{-\varepsilon
(g)}^{\left[ \varepsilon (g)\right] }}{\sqrt{E^{2}-m^{2}c^{4}}}\,\right]
^{\varepsilon (g)}C^{\left[ +\right] }  \label{24}
\end{equation}

\[
D^{\left[ -\right] }=i\,\sqrt{\frac{E-mc^{2}}{E+mc^{2}}\,}\,C^{\left[
+\right] }
\]

\noindent and the constant $C^{\left[ +\right] }$ is to be fixed by the
normalization condition. It is noticeable in (\ref{eq10}) a somewhat
left-right symmetry involving $\psi _{+}$ and $\psi _{-}$, such a symmetry
is not exact inasmuch as $C^{\left[ \pm \right] }\neq $ $D^{\left[ \mp
\right] }$. The upper and lower components of the spinor have the same
number of zeros, or nodes, and as a consequence of such a quasi-symmetry the
zeros of $\psi _{+}$ and $\psi _{-}$ exhibit, of course, an exact left-right
symmetry. In whatever manner Eq. (\ref{24}) of course, is invariant under
the change $E\rightarrow -E$ provided $|C^{\left[ \pm \right]
}|\leftrightarrow |D^{\left[ \mp \right] }|$, hence one can conclude that $%
|\psi _{+}(\pm x)|\leftrightarrow |\psi _{-}(\mp x)|$ in such a way that the
position probability density on the right side of the $x$-axis transforms
into the position probability density on the left side of the $x$-axis, and
vice versa. As for the combined transformations $g_{0}\rightarrow -g_{0}$
and $g\rightarrow -g$, one has $|C^{\left[ \pm \right] }|\leftrightarrow
|D^{\left[ \pm \right] }|$ with the proviso that $m\rightarrow -m$.
Recalling that the existence for bounded solutions demands that $\varepsilon
(g)g_{0}>1$, this last set of mathematical transformations allow us to
concentrate our attention on the case $g>0$. Figs. \ref{Fig2}-\ref{Fig4}
illustrate the behaviour of the upper and lower components of the Dirac
spinor, $|\psi _{+}|^{2}$ and $|\psi _{-}|^{2}$, and the position
probability density, $|\psi |^{2}=|\psi _{+}|^{2}+|\psi _{-}|^{2}$, as a
function of $x$, for the positive-energy solutions with $g=2$, $g_{0}=5$ and
$\lambda =\hbar /(mc)$. Figs. \ref{Fig5}-\ref{Fig7} do the same with $g=14$.
The normalization constant, $C^{\left[ +\right] }$, was fixed by numerical
computation. Comparison of these figures shows that $|\psi _{+}|$ is larger
than $|\psi _{-}|$ (for $E>0$) and that the fermion tends to avoid the
origin as $g$ increases. Furthermore, $|\psi |$ tends to concentrate at the
left (right) region when $E$ and $g$ have equal (different) signs. This
happens due to the discontinuity of the effective Morse-like potential at $%
x=0$, given by $V_{eff}^{\left[ \pm \right] }(0_{+})-V_{eff}^{\left[ \pm
\right] }(0_{-})=\pm $ $\hbar ^{2}g/(2\lambda ^{2}m)$. Note from all these
figures that the quantum number $n$ qualifies the number of nodes of $\psi
_{+}$ and $\psi _{-}$. Note also that the quasi-degeneracy of the second-
and third-excited states for $g=14$ reveals a near equality of the position
probability densities for those states, as should be expected.

A numerical calculation of the uncertainty in the position for the
ground-state solution (with $m=c=\hbar =1$ and $g_{0}=5$) furnishes $0.736$
and $0.469$ for $g=2$ and $14$, respectively. Here we have purposely shown
an odd fact for $g=14$. It seems that the uncertainty principle dies away
provided such a principle implies that it is impossible to localize a
particle into a region of space less than half of its Compton wavelength
(see, e.g., Ref. \cite{str}). This apparent contradiction can be remedied by
recurring to the concepts of effective mass and effective Compton
wavelength. Indeed, the third line of (\ref{15b}) suggests that we can
define the effective mass as
\begin{equation}
m_{eff}=\sqrt{m^{2}+\left( \frac{\hbar g_{0}}{2\lambda c}\right) ^{2}}
\label{35}
\end{equation}
Hence, the effective Compton wavelength can be defined as $\lambda
_{eff}=\hbar /(m_{eff}c)$ so that the minimum uncertainty consonant with the
uncertainty principle is given by $\lambda _{eff}/2$. Therefore, the
uncertainty in the position can shrink without limit as $|g_{0}|$ increases
or $\lambda $ decreases. It means that the localization of the fermion does
not require any minimum value as $|g_{0}|$ $\rightarrow \infty $ or $\lambda
\rightarrow 0$ in order to ensure the single-particle interpretation of the
Dirac equation.

\section{Conclusions}

We have succeed in searching for Dirac bounded solutions for neutral
fermions by considering a pseudoscalar screened Coulomb potential in 1+1
dimensions. The satisfactory completion of this task has been alleviated by
the methodology of effective potentials which has transmuted the question
for $E\neq \pm mc^{2}$ into Schr\"{o}dinger-like equations with effective
Morse-like potentials. There are no isolated solutions ($E=\pm mc^{2}$) for
this sort of potential. Nevertheless, as the coupling constant $g$ becomes
extremely large, i.e. $|g|\gg 1$, the energy levels corresponding to the
ground-state solution end up close to $\pm mc^{2}$.

The uniform background potential, beyond the screened Coulomb potential, is
a sine qua non condition for the existence of bound-state solutions.
Curiously enough, it plays an essential role not only to determinate the
attractiveness of the effective Morse-like potential but also for
establishing the band of allowed discrete Dirac eigenvalues. Furthermore,
due to the fact that there is no atmosphere for the production of
fermion-antifermion pairs, a neutron fermion embedded in this uniform
background acquires an effective mass which permits that it can be strictly
localized.

In addition to their intrinsic importance as new solution of a fundamental
equation in physics, the solutions obtained in this paper might be of
relevance to the confinement of neutral fermions in a four-dimensional
world. Furthermore, they render a contrast to the result and conclusions
found in \cite{mck}: there are bound-state solutions for neutral fermions
interacting by a pseudoscalar screened Coulomb potential in 1+1 dimensions,
notwithstanding the spinor is not an eigenfunction of the parity operator.

\bigskip\bigskip\bigskip

\noindent \textbf{Acknowledgments}

This work was supported in part by means of funds provided by CNPq and
FAPESP.

\newpage

\newpage

\newpage

\newpage

\newpage

\newpage

\newpage

\end{document}